# Feynman's Sum-over-Paths method applied in wave optics and for calculating the quantum probability current

J. Joerg

Munich, Germany
Email: josef.joerg@online.de

## Abstract

Based on the *Sum-over-Paths* approach of Richard Feynman, an integration method for calculating wave phase vectors is derived.

The diffraction and interference patterns of various slit masks can be calculated from such phase vectors. The results obtained match with those computed using the more complex Fresnel integrals.

Babinet's principle for phases is presented for the first time. As an example, the diffraction pattern of a double slit can also be calculated by subtracting the phase vectors of a smaller slit from those of a larger one. This method is also demonstrated by using a diffraction calculation tool when an obstacle is positioned behind a double slit.

The approach of phase vector integration is expanded to calculate the quantum probability current. Based on the Hamilton-Jacobi- and the stationary Schrödinger -equation, the quantum probability current is calculated and its continuity is shown.

The time-dependent differential equation of the probabilty current velocity is numerically solved. For a large number of integrations it is statistically shown that the distribution of the flux per area is consistent with the probability density profile.

The quantum probability current is also simulated for photons traversing an asymmetrical double slit and the special results are discussed.

## 1. Introduction

In the first chapters of his popular book *QED – The strange theory of light and matter* [1], Richard Feynman describes a method for calculating wave-like phenomena, e.g. reflection or diffraction at the grating. This method is known as the *Sum-over-Paths* model, which means that light or a wave-like particle seems to take all possible paths moving from point A to B.

Richard Feynman draws out the time dependency of this method and later on - understandable for laypeople - the famous Feynman diagrams, the foundation of *Quantum Electrodynamics*. The QED theory considers many interaction types with virtual particles and is one of the most commonly-tested theories. As an example, the anomalous magnetic moment of an electron can be measured with extremely high precision. The deviation between experiment and the predicted value by the QED is better than one part in a trillion.

In this paper we will use the *Sum-over-Paths* method for the stationary case, neglecting any time dependency. Additionally we will not consider the dispersion of wave-like particles in time. The method may be compared with the *Stationary Phase* model in wave optics, where phases simply depend on the space coordinates. The stationary method will be sufficient for the calculations and simulations to be worked out here.

## 2. *Sum-over-Paths* Model Applied on a Single Slit

We start with the summation over all possible paths with a simple case, where a slit is put between the particle source and the two-dimensional measurement point at coordinates (x, y).

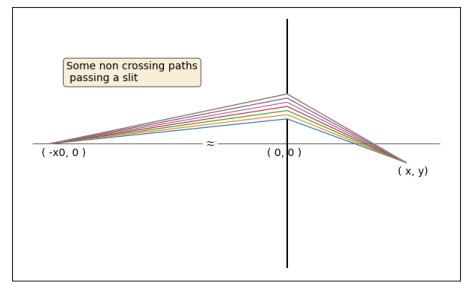

Fig. 1: Paths used for the *Sum-over-Paths* integration

In Fig. 1, the chosen coordinate system and some of the possible paths are shown.

The slit is slightly shifted with $D$ in y-direction, which is later used to calculate interference patterns.
The figure is not true to scale, and the following requirements according to wave optics have to be considered:

- light or particle source distance $x_0 \gg D$ and
  $x_0 \gg$ slit width *slitw*

- distance $r = (x^2 + y^2)^{1/2} \gg D$, $r \gg$ *slitw*,
  $r \gg$ wavelength $\lambda$

According to Fig. 1, the length of a path can be calculated by simply using Pythagoras (variable $d$ varies within the slit width):

$$l_{path} = (x_0^2 + d^2)^{1/2} + (x^2 + (y - d)^2)^{1/2}$$

The components in x,y direction of a phase vector for this path results in:

$$p_x(x, y) = \cos(k * l_{path})$$
$$p_y(x, y) = \sin(k * l_{path})$$

with wavelength $\lambda$ and wave number $k$

$$k = 2\pi/\lambda$$

This can also be written in the complex plane:

$$p = \cos(k * l_{path}) + i * \sin(k * l_{path})$$

Simply to make the phase vector with the components ($p_x$, $p_y$) more illustrative:
The length of the path is divided by the wavelength $\lambda$, whereby only the remainder $r$ is valuable for the trigonometric functions. If $r$ has a value equal to 0 or $\lambda$, the phase vector points to +x direction, for $r$ equal $\lambda/4$, the phase vector points to +y direction, for $r$ equal $\lambda/2$ it points to the -x direction and so on. The angle $\varphi$ between x-direction and phase vector takes values between 0 and $2\pi$ with counter-clockwise direction.

Summing up the phase vectors for N paths results in:

$$p_x = \sum_{i=0}^{N} p_{x\,i} \qquad p_y = \sum_{i=0}^{N} p_{y\,i}$$

For $N \to \infty$ the sums can be replaced by the following integrals:

$$p_x = \int_{y_D=D-slw/2}^{y_D=D+slw/2} \cos(k * p(x, y, y_D)) \, dy_D$$
$$p_y = \int_{y_D=D-slw/2}^{y_D=D+slw/2} \sin(k * p(x, y, y_D)) \, dy_D \quad (1)$$

with:
$$p(x, y, y_D) = (x_0^2 + y_D^2)^{1/2} + (x^2 + (y - y_D)^2)^{1/2}$$

$p(x, y, y_D)$ is simply the addition of the two paths for $x \leq 0$ and $x > 0$. The parameter $D$ is the distance to the middle of the slit (the usually used lattice constant equals $G = 2 * D$).

These integrals cannot be solved analytically, but rather numerically.

For the computational work, the programs have been implemented in *Python* language. *PyDev* was used as development environment. It is an IDE that is integrated in the *Eclipse Software Development Kit*. The hardware was a common notebook.
The integration of the equations (1) is performed via the scientific *Python* package *scipy* (function *integrate.quad*).

After integrating the phase vector in equation (1), we can obtain the intensity in optical meaning, which in our case also corresponds to the probability density function in quantum mechanics. The intensity is proportional to the squared length (or amplitude) of the sum-over-paths phase vector:

$$intensity \propto p_x * p_x + p_y * p_y$$

We should note here that for this method no specific characteristics of light (e.g. E- / B-fields) are taken into account. The *de Broglie* wavelength for a moving particle can also be used.
The intensity value is still not normalized. The number of particles per angle element $d\varphi$ must remain constant. Since the arc-length in a xy-plane increases with $r * d\varphi$, the intensity has to decrease proportional $1/r$.

In order to be certain that the intensity value above does not contain another $r$-dependency, curve integrals along a semicircle for various distances $r$ were calculated. These integrals for the intensity expression above remained constant within a wide range of realistic numerical stable distances.
The distance, where the intensity should be 1 was defined as $R_0 = slit\text{-}width/2$.

Therefore we obtain for the normalized intensity:

$$intensity_{norm}(x, y) = R_0/r * (p_x * p_x + p_y * p_y)$$

In Fig. 2 the diffraction patterns of a single slit versus the y-axis for some distances x are shown, according the results of the equations (1) with D = 0.

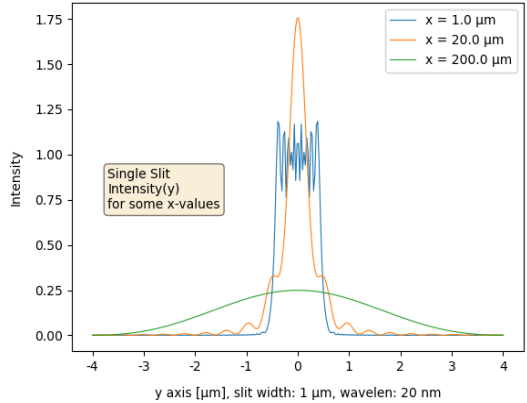

Fig. 2: Intensity distribution of a single slit for various distances x

The wavelength 20 nm was chosen according to the *de Broglie* wavelength of ultra-cold Ne-atoms with a velocity of 1 m/s. This refers to the publication of Michel Gondran et al. [2], who conducted comprehensive time-dependent simulations and evaluations on the experimental data of Fujio Shimizu et al. The results obtained by Gondran et al. are in excellent accordance with the results obtained via equation (1), despite the time dependency there and the time independence here.

The results obtained for a single slit can easily be extended to an interfering double slit. If we take the results of (1) at the y-distances -D, +D, a simple addition of the phase vectors and squaring will give the interference patterns. For non-interfering single slits, the intensities have to be added together.

$$p_{x1}, p_{y1} = single\ slit(-D)$$
$$p_{x2}, p_{y2} = single\ slit(+D)$$

$$double\ slit\ p_x, p_y = p_{x1} + p_{x2},\ p_{y1} + p_{y2}$$

In Fig. 3 the well-known interference patterns of a double slit together with the sum of intensities of two (non-interfering) single slits are shown.

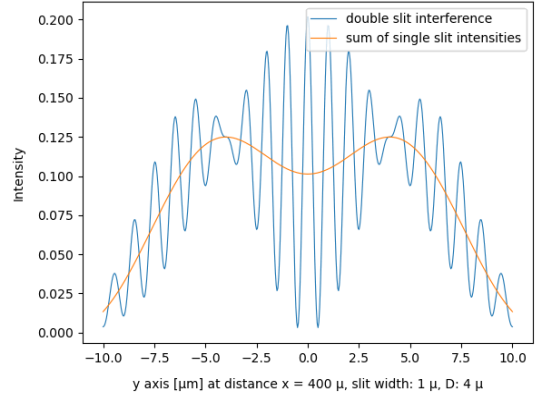

Fig. 3: Intensity distribution of a double slit at distance x = 400 μm together with the sum of intensities of two single slits

In Fig. 4 the intensity increase and decline for a double slit at y=0 is shown. It should be noted here that at a high resolution, the oscillations of the intensity with wavelength λ would be seen.

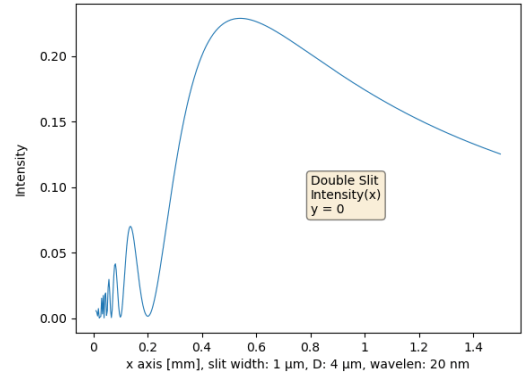

Fig. 4: Intensity increase and decline in x-direction of a double slit for y = 0

## 3. Comparison with Fresnel Diffraction

The results obtained by the integration method (1) can be compared with the Fresnel diffraction in wave optics. A very good overview and a few application examples are given in [3]. Within these theories, much more mathematical effort is needed, compared with the *Sum-over-Paths* method used in (1).

The use of Fresnel integrals for a single slit results in the following formulas, which are excellently explained in [3]. Our paper makes the novel contribution that the phase vectors are calculated and used. For computing the Fresnel integrals, the *Python* library *scipy.special* has been included.

*Note*: the indices $U / L$ mean the upper / lower edge of the slit, *slitw2* means slit-width/2, $C$ and $S$ are the Fresnel integrals, evaluated at a given point (x, y).

$$v_u = (2 / (\lambda * x))^{1/2} * (y + slitw2)$$
$$v_l = (2 / (\lambda * x))^{1/2} * (y - slitw2)$$
$$S_u, C_u, S_l, C_l = fresnel(v_u), fresnel(v_l)$$
$$p_x = (C_u - C_l) / \sqrt{2} \qquad p_y = (S_u - S_l) / \sqrt{2} \qquad (2)$$
$$intensity = p_x * p_x + p_y * p_y$$

For calculating the intensities for a double slit, we simply have to shift the reduced variables $v_u$, $v_l$ by +D, -D in y-direction (same as in the chapter before), add the resulting phase vectors and square the sum to obtain the corresponding interference patterns.

In the following diagram Fig. 5, the comparison of the two methods (1), (2) is shown. There is an excellent accordance within the numerical precision.

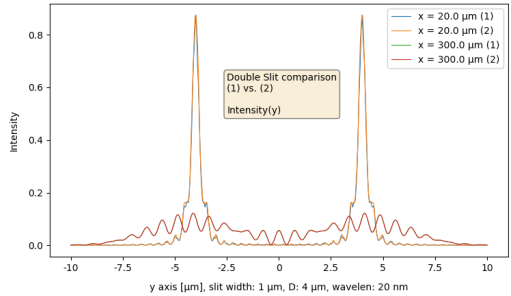

Fig. 5: Double slit intensity comparison of method (1) with Fresnel diffraction (2)

The excellent matching between the *Sum-over-Paths* method (1) and the Fresnel method (2) has to be further investigated. The two methods originate in completely different approaches. The Fresnel method has the advantage that the distance dependency is already considered by the integrals.

The interference patterns were also be compared with the Fraunhofer approach and mach exactly for larger distances. At smaller values of x (for our example x < 1mm), the Fraunhofer approach breaks down compared with the *Sum-over-Paths* method (1) and the Fresnel method (2).

It is remarkable in our perception that in most publications in the context of Fresnel diffraction, the intensities play the main role, but not the phases. Indeed, intensities are the only physical sizes that can be measured. However, phases are quite important in other contexts of wave optics and quantum mechanics. We will use Fresnel phases for Babinet's principle expanded for phases and later on for calculating the quantum probability current.

The mathematical scope of equation (1) corresponds to the requirements of wave optics in the Fresnel zone and larger regions. There are heavy phase oscillations at distances smaller than 20 wavelengths and standing waves in y-direction at x = 0. The evaluations in this area are not needed for the calculations in this paper.

The numerical integration of equation (1) is quite fast, but within *Python* the calculation of the Fresnel integrals seems to be implemented in a very fast algorithm. Filling a matrix of 401,401 elements takes less than 2 minutes on a common notebook. The intensity matrix is shown in Fig.6.

Notice that in this figure the intensities at greater distances are slightly enlarged, otherwise the intensity color would fade out too rapidly.

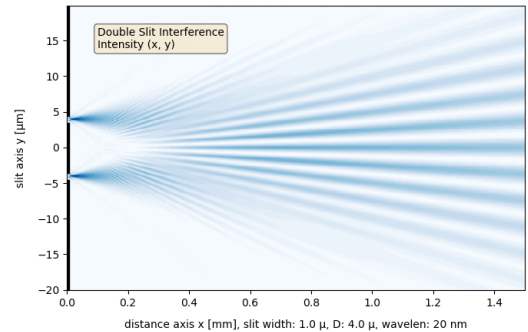

Fig. 6: Interference pattern of a double slit

## 4. Babinet's Principle Expanded for Phases

To obtain the diffraction of an obstacle within a plain wave, it should be the same as subtracting the diffraction of an opening in a mask from the plain wave. This is known as Babinet's principle.

We conducted extensive online research, if Babinet's principle can be applied to phase vectors. But we have found only that this principle is used for E-fields and intensities.
In this paper, we use Babinet's principle expanded for phases for the first time. For the example of a double slit, the following tasks have to be performed using the equations (2) for centered single slits
(*slw* means the slit width of the double slit):

- first calculate $p_{x1}$, $p_{y1}$ for slit width:
  (2*D + slw)

- second calculate $p_{x2}$, $p_{y2}$ for slit width:
  (2*D – slw)

- third subtract
  the phase vector of slit2 from that of slit1:
  $p_x$, $p_y$ = $p_{x1}$, $p_{y1}$ - $p_{x2}$, $p_{y2}$

This means that according to Babinet's principle we are subtracting a smaller slit mask from a larger one. In our case, we subtract a phase vector from another one.
Calculating the intensities for $p_x$, $p_y$ results in intensities exactly to those obtained from equation (2) for the double slit. If we simply subtract the intensities from each other, we would derive an incorrect result.
For the sake of completeness the comparison is shown in Fig. 7.

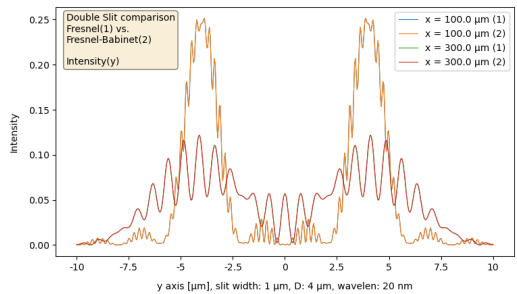

Fig. 7: Double slit comparison Fresnel vs. Fresnel-Babinet

The method used for a double slit can easily expanded for a triple slit by simply adding the phase components of a single slit at y = 0.

The principle of subtracting phase vectors from each other can also be used for x-distances > 0.

To demonstrate this, we used the *Python* library *Diffractio*. In this library, algorithms for diffraction and interference optics are implemented. By graphically placing geometrical rectangles in an appropriate way, all necessary masks like slits, double slits or obstacles can be modeled. For the evaluations, the *Beam propagation method (BPM)* algorithm has been taken.
For our example, we used green laser light (wavelength 520 nm) as the light source. The following masks have been implemented:

- double slit (slit width = 100 μm, D = 200 μm)
- obstacle (width = 100 μm, x=20 mm, y-shift: 100 μm)
- aperture (diameter, position: same as the obstacle)
- measurement distance: 70 mm

First, we place only the obstacle at x=20 mm behind the double slit. The original symmetrical double slit pattern distorted by the obstacle is shown in Fig. 8. Next, the obstacle is replaced by an aperture placed behind the double slit at the same position. Since most parts of the light are covered in this case, there is only a low intensity measured (see Fig. 8).
We now use Babinet's principle expanded for phases:

$$\vec{p_{obstacle}}(x,y) = \vec{p_{dslit}}(x,y) - \vec{p_{aperture}}(x,y)$$

Note that within *Diffractio*, complex numbers are used for the phase vectors. As already highlighted in chap. 1, this is identical to our notation, where the real part corresponds to the $p_x$ phase vector component, and the imaginary part corresponds to the $p_y$ component.

Using the *Diffractio* phase vectors (after intermediate storage and read in again) along the measurement axis, we obtain the intensities shown in Fig. 8. They match exactly with those intensities previously obtained for the obstacle alone.

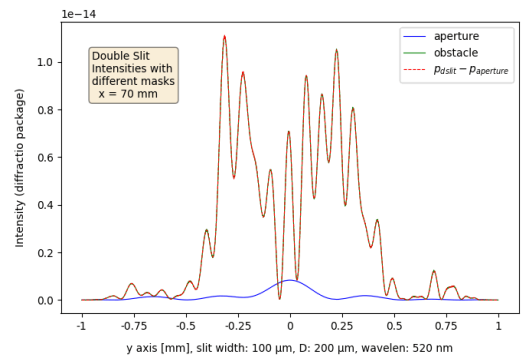

Fig. 8: Double Slit intensities with different masks

# 5. Calculating the Quantum Probability Current

There are many publications describing the relationship between the Hamilton-Jacobi- and the stationary Schrödinger -equation. An overview is provided in [4].
In summary, we use only the following two formulas:

$$\Psi = A\,e^{\,i\,S/\hbar} = A\,e^{\,i\,\varphi}$$

with the wave function $\Psi$, the Hamilton's principle function $S$ (or action function), the phase $\varphi$ and the reduced Planck's constant $\hbar$. The Sum-over-Paths phases may also be interpreted as *Least-Action*. Richard Feynman often highlighted this context; for example in *The Feynman Lectures on Physics* [5]

The second formula used in the following chapters describes the p*robability current j* (see e.g. [6]):

$$\vec{j} = \rho * \vec{v} = \rho/m * \vec{\nabla} S = \rho * \hbar/m * \vec{\nabla}\varphi \quad (3)$$

with the probability density $\rho(x,y)$, the particle velocity $v$ (or group velocity of the wave), the particle mass $m$ and the gradient on the *S*-function (or phase $\varphi$).
We already used the x- and y-components of the summed-up phase vector in equation (1). For their integration, we use now the faster method via the Fresnel integrals equation (2). The phase itself can be calculated by:

$$phase = \arctan\left(\,p_y/p_x\,\right)$$

It is a rapidly oscillating function in the range [ $-\pi$, $+\pi$ ]. The oscillations in y-direction are shown in Fig. 9 for a double slit. For the wavelength of 20 nm, the distance x = 150 μm is in a moderate range. For smaller x, the oscillations rapidly increase.

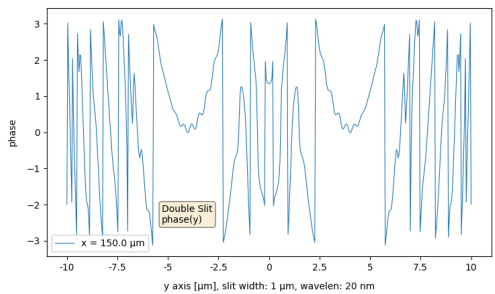

Fig. 9: Oscillating phase in the range [ $-\pi$, $+\pi$ ]

In order to create gradients on the phase angle function, it has to be expanded. If the phase difference of two neighbored y-points is greater/less than $+\pi/-\pi$, then $+/- 2\pi$ has to be added. This is shown in Fig. 10 for various values of x-values.

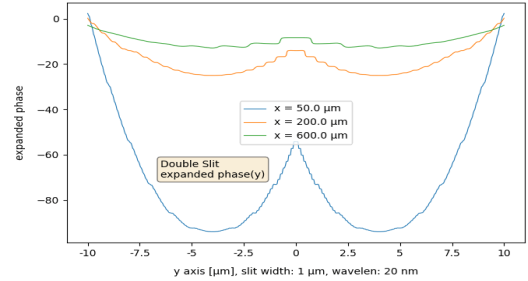

Fig. 10: Expanded phase values for various x-values

In this figure, the turning points can already be seen, which will result in minimums / maximums for the gradients (equation 3).

According to equation (3), the gradients of the phase in x, y direction multiplied by $\hbar/m$ can be identified as the velocities of the probability flux. The velocities are shown in Fig. 11 at some x and y distances for a double slit in x- and y-directions.

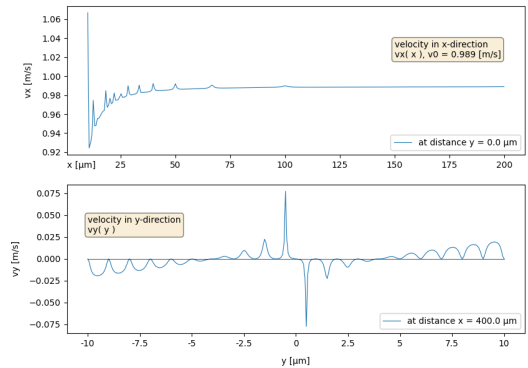

Fig. 11: Some velocities of the probability current for a double slit

The x-velocity for larger x-distances tends to the starting velocity v0 of a particle, in our case 0.989 m/s (not 1 m/s, as used by Gondran et al. in [2], because instead of 19.8 nm, 20 nm has been taken for the wavelength). The velocities in y-direction decrease/increase for negative/positive *y* and have minima and maxima, which result in characteristic deflections in the pathlines shown in the next chapter. The mean velocities in y-direction are smaller by a factor of 10 to 50 than in x-direction.

For the numerical stability of differentiation, the step size may neither be too small nor too large (see e.g. [7]). In our calculations, we obtained stable results in a wide range of distances and wave lengths with
step size $h_x$ = wavelength / 2000, and
step size $h_y$ = wavelength / 20.
Again, we have to expand the phase, if there is a phase jump between (x – h) and (x + h).

**Note:** If we use in equation (3) for the absolute value of **v** approximately the initial velocity v0, the absolute value of **j** is proportional to the probability density ρ.

The calculation of the velocities $v_y$ in Fig. 11 can also be used in another context.
Kocsis et al. [12] experimentally performed weak measurements on the transverse momentum of photons in a two-slit interferometer. In a commenting paper Bliokh [13] argued that the measurements can also be interpreted as the transverse Pointing vector. In any case the wave vector component $k_y$ can be calculated by just taking the derivative of phase φ with respect to y (or equivalent $k_y = v_y * m / \hbar$).
In Fig. 12 the wave vector component $k_y$ along the y-coordinate is shown. The profile can directly be compared with the experimental data ref. [12] Fig. 2D.

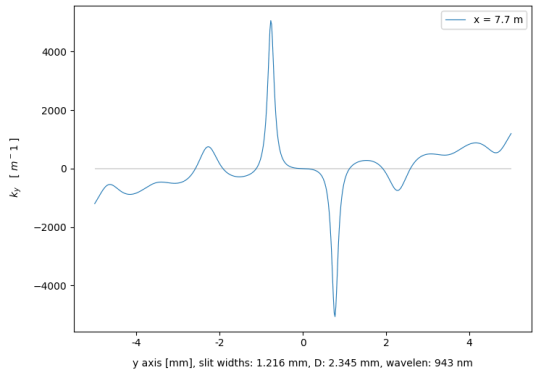

Fig. 12: Transverse wave vector component $k_y$
(for the used settings see ref. [12])

## 6. Integrating the Differential-Equation

From equation (3), an ordinary differential equation system in time can be derived:

$$d/dt\ \vec{x} = \vec{v}(x,y) \quad (4)$$

The advantage of solving this ODE lies in the fact that if we take the probability flux at a starting point (x, y), we can observe the evolution of the flux to new points (x1, y1).

For solving the ordinary differential equation (4), the Python solver *solve_ivp* (package *scipy.integrate*) was used [9]. The method within the solver was adjusted to *LSODA* (Adams/BDF method with automatic stiffness detection and switching). The relative / absolute tolerances *rtol, atol* were set to *1e-10*.

Gondran et al. [2] solved the differential equation above for time-dependent wave packets, which result in the well-known *Bohmian trajectories* [6]. For the simulations shown in Fig. 13, we took the same experimental data as in the paper of Gondran. Fig. 13 can be directly compared with Fig. 10 in [2]. The starting points of the paths were statistically chosen according to the probability (e.g. Fig. 5) at 20 μm in a symmetric way. An important characteristic of the interfering pathlines is that they do not cross each other. Therefore they also do not cross the x-axis at the values y = 0. In Fig. 13, we again see the deflections of the pathlines across the minima, resulting from the minima / maxima of the velocities in y-direction (see Fig. 11). For a few number, the non-interfering, crossing paths are also shown in Fig. 13 in color *magenta*.

It is quite interesting that the main maximum at y = 0 at larger x-distances is summed up by the deflected probability current pathlines at shorter distances.

It should be noted that the expansion of the pathlines in y-direction is just very small. The aspect ratio of x:y in Fig. 13 is more than 140:1. Moreover, the deflection curvature in y-direction is very small. The deflections e.g. in the middle of Fig. 13 take up approximately 0.6 μm in y-direction, but 100 μm in x-direction.

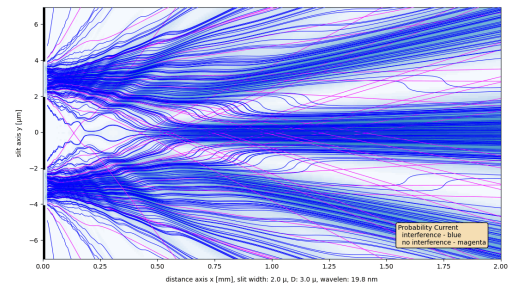

Fig. 13: Probability density current pathlines for a double slit
(in *magenta* color: non-interfering pathlines)

As already mentioned the probability current pathlines in Fig. 13 are already known as *Bohmian trajectories*. There are many discussions about the *de Broglie-Bohm* interpretation, which will not be further handled here. In most cases, this interpretation does not provide any new insights above the common quantum mechanics. On the other hand, the *de Broglie-Bohm* interpretation can clarify some peculiar issues like *Wheeler's delayed-choice experiment* [10] and other mysteries on similar topics.

In this paper we will avoid the misleading term *trajectories*. For conditions where Heisenberg's uncertainty principle has to be taken into account, there will be no observable trajectory of a particle.

Therefore we will simply use the term *probability current* (or *flux*) and its evolution in time and coordinates (x, y) according to the equations (3, 4). Nevertheless, the probability current is closely related to the probability of a current of particles.

## 7. Continuity of the Probability Current

The time-dependent continuity equation in quantum mechanics reads as:

$$\partial \rho / \partial t + \text{div}\, \vec{j} = 0$$

For the stationary case, it follows that the divergence of the probability current j has to be zero. This is equivalent that the probability current (3) passing through a unit area has to be constant in any direction. The effective probability current in y-direction for an area between x, x+dx turns out to be zero, because the probability density ρ in equation (3) is symmetrical to the x-axis, and the velocity $v_y$ is anti-symmetric (see $v_y$ in Fig. 11).

For the probability current in x-direction, we have to integrate across the area in y-direction:

$$j_x(x) = \int_{y1=y_{mostLowerPath}}^{y2=y_{mostUpperPath}} \rho(x,y) * v_x(x,y) \, dy \quad (5)$$

For the integration limits, we simply use the most lower and the most upper pathlines. The integration is again performed via the scientific *Python* package *scipy* with the function *integrate.quad*.

As an example the integration was performed for the case of a double slit (see Fig. 14). In this example, the paths were initialized for the lower slit ($y_0$=-4.0 μm) at:

x = 20 μm, y >= -4.15 μm <= -3.85 μm

The pathlines are shown in alternating colors to better illustrate the streaming characteristics of the probability current. Again we see the deflections corresponding to the minima / maxima of $v_y$.

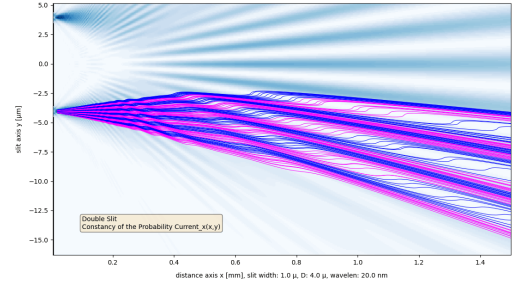

Fig. 14: Conservation of the probability current in x-direction

The constancy of (5) could be verified in the region of 50 μm - 1.5mm with a mean relative standard deviation less than 2e-5.

For another example similar to Fig. 14, the region expands between 0.1 mm and 10 cm, the distance between the most lower and most upper path at the end of the probability flux is about 120 times greater than at the start. The mean relative standard deviation of the integrals (5) is less than 1e-5.

## 8. Probability Current per Unit Area

In order to calculate the probability current crossing an area element (in our case a 1-dim Δy) at a distance x, a large number of statistically initialized differential equations (4) have to be integrated. The initialization was again conducted at x = 20 μm with a y-value density distribution best matching the probability density at this distance (see e.g. chap.3, Fig. 5). For our example, there is nearly no difference of the double slit probability density versus the distribution of two single slits at x = 20 μm.

The initialization and integration were performed for about 10,000 paths. The integration result of a pathline is stored by the solver *Python solve_ivp* in a file with typically 600 – 1100 *x-y* coordinate segments. To obtain the pathlines, crossing a defined *x*-distance, the segments of all pathlines have to be analyzed. If there is a crossing of a segment, the crossing point (*x,y*) has to be calculated. Counting these points in an interval Δy, we obtain a measure of the probability flux, crossing this element.

In Fig. 15 and 16, histograms of the number of crossings per Δy at a distance x are shown. Additionally, the scaled probability density ρ is drawn (adapted to the maximum number of pathlines in the histogram). It is shown again, that there is no crossing of the x-axis at y = 0. The different colors mark the pathlines starting from the left or right side y<=0,y > 0.

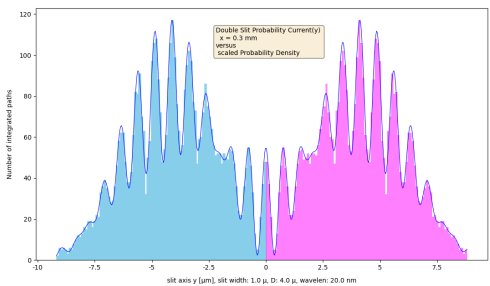

Fig. 15 Histogram of the probability current

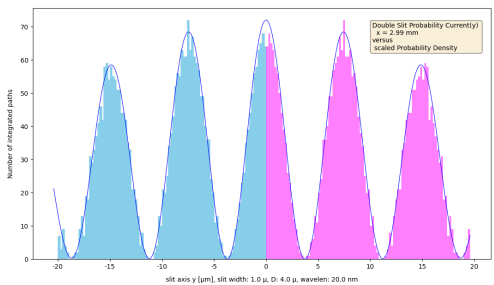

Fig. 16: Histogram of the probability current

The diagrams show that within statistical deviations, there is no principal difference between the function profile of the probability current and the probability density ρ. High probability densities are equivalent to a high density of the probability current pathlines, while lower probability densities correspond to lower current densities (see Fig. 14). It should be emphasized again, that we simply used the probability density to initialize the differential equations at a relatively small distance. The probability profile automatically results by calculating the probability current velocities and integrating them over time.

The probability current more accentuates the flow of particles. But it should be emphasized that it is not an observable in quantum mechanical sense.

In the view of standard quantum mechanics each measurement causes the collapse of the wave function. The measurement results are distributed according to the probability density (calculated by the squared wave function).

Considering the above, a double slit experiment with particles can be interpreted in the following manner.
We start with the time-dependent wave function of a free particle, which can be described by a wave packet function with a corresponding *de Broglie* wave length. The wave packet spreads out in space and momentum by time. The phases of the wave packet are distorted by the two slits. According to the respective probability, a particle traverses one of the slits, where the phases of the distorted wave packet determine the further evolution of the probability current.

As already mentioned, the time-dependent simulations on this topic were conducted in detail by Michel Gondran et al. [2]. The further analysis of the probability current per unit area was not performed in their publication.

It is quite remarkable that the results in [2] can be obtained simply by evaluating the stationary case figured out in our paper. Moreover, the complex interactions close to, or within the slits are not relevant for the results.

It also should be mentioned, that the probability current makes no sense for bound states. For example, for atoms, the probability current calculates to zero.

## 9. Asymmetrical Probability Current

For symmetric double slits, it is not possible to follow the probability current from smaller to larger distances. The pathlines fade out to neighbored minima or maxima (see Figures 13 and 14). Various simulations showed an interesting behavior for asymmetric double slits.

We now use a wavelength of 520 nm, equivalent to the light of a green laser. In Fig. 17, the intensities for an asymmetric double slit for various distances are shown. The lower slit has a diameter of 100 μm, and the upper slit measures 50 μm. The intensities were calculated via the Fresnel integrals equation (2).

Note, that the micrometer scaling in the previous chapters now turns to millimeter sizes.

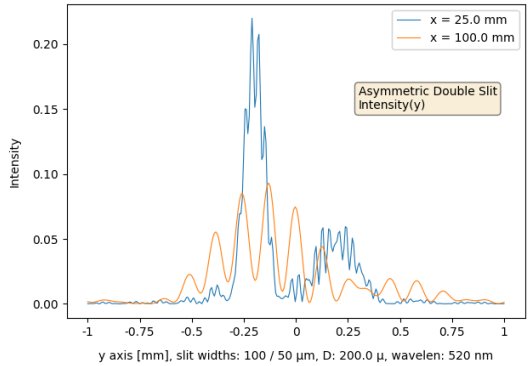

Fig. 17: Intensity distribution of an asymmetric double slit for various distances x

For calculating the proportionality factor ℏ/m in equation (3), we now have to consider the mass of a photon with wavelength 520 nm. But for light, it is more convenient to use the wave vector (see Fig. 12).

For the integration of the differential-equation (4), the simulations were initialized according to the probability density distribution at 2 mm and end up at 90 mm. There were about 12,000 simulations, which are depicted thinned out in Fig. 18.

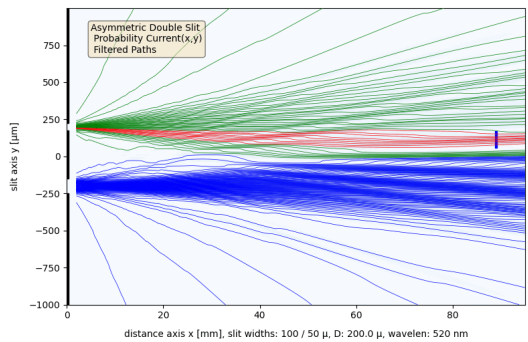

Fig. 18: Probability current for an asymmetric double slit (the paths are thinned out)

The paths in *red* color, which build up the first maximum at x = 90 mm, y ~ 110 μm, were filtered out. This maximum is not a mixture of very different pathlines but rather originates from a small area at the start of the simulation.

The corresponding histogram for the simulations is shown in Fig. 19.

The probability density in Fig. 19 was also verified via the *Python* package *Diffractio* [11], already described in chap. 4. There, it could be shown that the first maximum vanishes if an obstacle of appropriate size is placed at 89 mm.

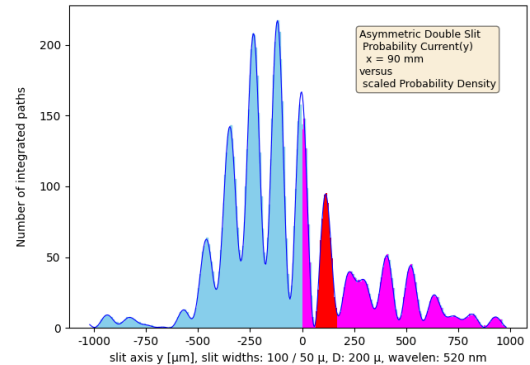

Fig. 19: Histogram of the probability current

It would be quite interesting to simulate the probability current paths, when an obstacle (or an aperture) with appropriate size is placed in the course of the paths. This could not be performed by the package *Diffractio* because it is grid-based. The grid could not be adjusted sufficiently small to obtain the velocities of the probability current (gradients of the phase vectors) with stable results.

We mention that the probability current reminds in some way of fluid dynamics, where the flux per area and its evolution in space and time is also analyzed. This view within quantum mechanics more accentuates the particle-like view. Nevertheless we obtain identical results for the wave-like view. In standard quantum mechanics, any measurement causes the collapse of the wave function and we get as result the probability density distribution.